\documentclass[aps,pra,english,groupedaddress, 11pt]{revtex4-2}
\usepackage{graphicx}
\usepackage{amsmath}
\usepackage{amssymb}
\usepackage[font=small,labelfont=bf]{caption}
\usepackage{amsfonts}
\usepackage[margin=2cm]{geometry}
\usepackage{mathtools}
\usepackage{multirow}
\usepackage{array} 
\usepackage{subcaption}
\usepackage[labelfont=bf, labelsep=period, font=small]{caption}
\usepackage{dcolumn}
\usepackage{seqsplit}
\usepackage{relsize}
\usepackage{mathrsfs}
\usepackage{calrsfs}
\usepackage{graphicx}
\usepackage{booktabs}
\usepackage{physics}
\usepackage[usenames,dvipsnames]{color}
\usepackage{xcolor}
\usepackage{hyperref}
\hypersetup{
    colorlinks=true,
    linkcolor=blue,        
    citecolor=teal,        
    filecolor=magenta,     
    urlcolor=violet,       
}

\begin{document}
\title{
Optimization of cooling power of
a thermoelectric refrigerator: A unified approach}
\author{Rajeshree Chakraborty}
	\email[e-mail: ]{mp18011@iisermohali.ac.in}
	\author{Ramandeep S. Johal} 
	\email[e-mail: ]{rsjohal@iisermohali.ac.in}
	\affiliation{Department of Physical Sciences,\\
		Indian Institute of Science Education and Research Mohali,\\
		Sector 81, S.A.S. Nagar,\\
		Manauli PO 140306, Punjab, India.}
		
\begin{abstract}
    We analyze the steady-state formalism for optimizing the cooling power of a thermoelectric refrigerator (TER), unifying the endoreversible and exoreversible approximations within one framework. Although the cooling power is non-optimizable within the endoreversible model based on Newtonian heat-transfer law, we show that the issue can be circumvented in the near-reversible regime 
    where the external thermal conductances
    are large, but finite. We extend this analysis to optimize 
    the cooling power 
    in the presence of both internal and external irreversibilities and derive 
    a closed-form expression for the 
    coefficient of performance (COP) 
    that depends on the thermoelectric figure of merit  and the ratio of internal to external thermal conductances. The model  
    reproduces the endoreversible and the exoreversible limits as special cases. We conclude that for small temperature differences, the combined irreversibilities reduce the COP to values below 1/2, which aligns with the observed performance of the single-stage TER, and can provide realistic estimates for the COP.
\end{abstract}
\maketitle
\section{Introduction}
Thermoelectric cooling systems promise a viable alternative to vapor compression refrigeration systems owing to their quiet operation, compact size, absence of moving parts, and eco-friendliness. 
This makes the thermodynamic analysis of these systems as
an essential guide for the design and implementation of 
theoretical models. All such devices operate in 
 irreversible regime leading to dissipation or
entropy generation. 
In general, a reversible heat engine and refrigerator 
can be regarded as exact duals of each other, while their
irreversible counterparts do not show this duality. This issue has bearing on the choice
of a suitable figure of merit to optimize the performance of an 
irreversible refrigerator. Thus, whereas
power output of a finite-time heat engine 
is a natural target function 
to maximize, there is no clear consensus on an appropriate
function for the finite-time model of a refrigerator.
Various proposals have been made in the past and their 
limitations have been discussed \cite{Velasco1997,Velasco_irrvrefrigerator1997, Izumida_2013,Long2015,Tomas2013,Yan1990}. 

Two simplifying assumptions play a useful role in 
these finite-time thermodynamic models of irreversible machines. The endoreversible approximation \cite{Curzon1975}
considers the only source of irreversibility to be
the non-ideal thermal contacts between the reservoirs
and the working medium, assuming the internal processes
in the latter to be reversible. 
The exoreversible approximation \cite{Schmiedl2008} deals with 
the complementary situation of only
the internal irreversibility and the thermal contacts with the environment are regarded ideal. 
These approximate models have been considered 
for both discrete stroke cycles as well as in 
steady-state regime such as for thermoelectric devices \cite{Agrawal1997,Pedersen2007,Kaur2022}. 
While the endoreversible heat engines can be  optimized 
with different target functions \cite{Angulo1991,Chen1994,Medina2019}, the corresponding approach for 
endoreversible refrigerators has been less straightforward. 

Here, we consider the concrete example of the performance of a thermoelectric refrigerator (TER).
In Ref. \cite{Apertet2013}, it was argued that the cooling power of a TER in the endoreversible approximation cannot be optimized, in contrast to the case of an exoreversible approximation.  
The endoreversible model requires a specific law of 
heat transfer that is mediated by a heat exchanger of a finite thermal conductivity  between the reservoir and the working medium. 
Usually, for small temperature gradients,
this law is taken in the Newtonion form that regards the heat flux as 
 proportional to 
the temperature difference across the heat exchanger. Further,   in the engine mode, the model yields analytically tractable solutions for 
the maximum power output. This also 
leads to the famous Curzon-Ahlborn formula \cite{Curzon1975} for the efficiency
at maximum power, which is independent of any model 
parameters.  On the other hand, for the refrigerator mode, 
the cooling power seems a natural choice for the target function.
Here, assuming Newton's law for heat exchange with the reservoirs 
does not lead to an optimum w.r.t. to the 
cooling power \cite{Agrawal1990}. 
This difficulty motivated the introduction 
of alternate target functions for endoreversible refrigerator 
\cite{Yan1990, Mahler2010,Tomas2012, Luo2013}. 

On the other hand, the exoreversible 
refrigerator is based on the linear-irreversible
framework \cite{Apertet2013}.
Although Newton's law also presumes small temperature
gradients, it is a phenomenological law 
and is not well grounded in a theoretical framework.
Thus, a natural question is whether the endoreversible model
based on Newton's law lies within the linear response regime. 
We base our analysis on a steady state model 
of TER. 
In this paper, we show that operating close to the reversible limit (with thermal contacts deviating slightly from 
the ideal limit), the cooling power of an endoreversible TER can in fact be maximized.
\subsection{Model of TER}
 \begin{figure}[h]
     \centering
     \includegraphics[width=0.3\linewidth]{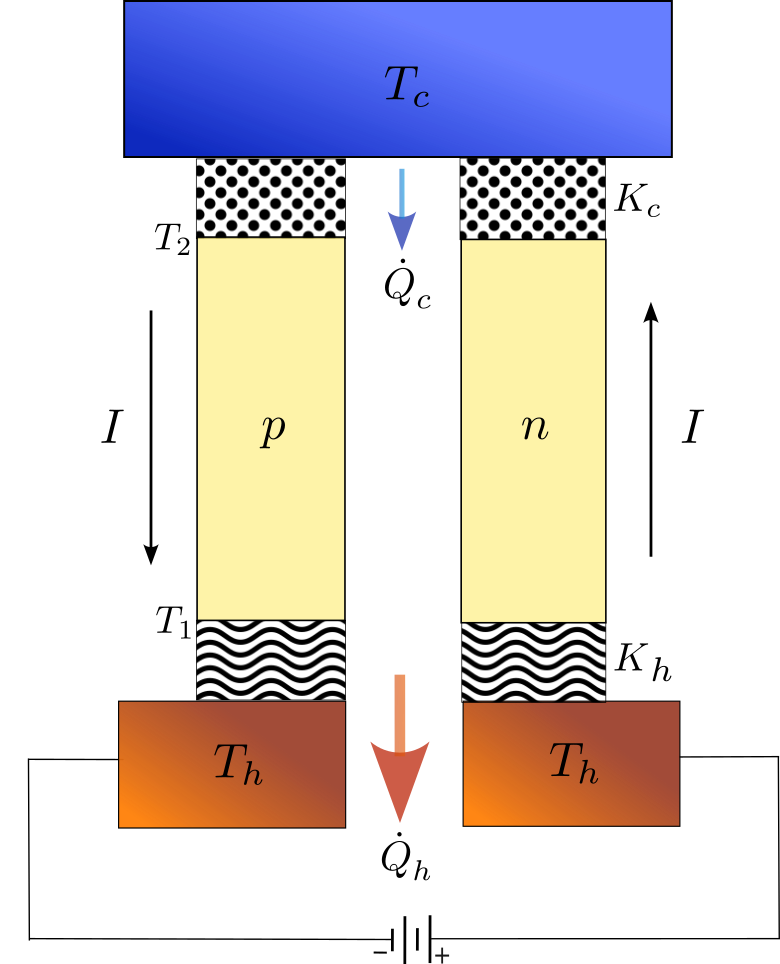}
     \caption{Schematic of a TER with $n$- and $p$-type legs connected electrically in series and thermally in parallel. $T_{1}$ and $T_{2}$ denote the local temperatures at the hot and cold ends of the thermoelectric element, respectively. $K_{h}$ and $K_{c}$ are the finite thermal conductances of the hot and cold exchangers, respectively. In the present paper, we consider symmetric conductances, 
     $K_h = K_c=K$.}
     \label{TERmodel}
 \end{figure}
A TER consists of a thermoelectric element driven electrically by an external power source and coupled thermally to two reservoirs at temperatures $T_h$ and $T_c$ ($T_h > T_c$). When an electric current $I$ flows through the device, heat is absorbed at the cold junction and rejected at the hot junction via the Peltier effect, while irreversible Joule heating and heat conduction occur within the thermoelectric material.

In a realistic device, external electrical or thermal  contacts are imperfect. 
In this paper, we ignore the electrical contact resistances \cite{Parrot1961} 
and focus only on the thermal contact resistances. As a result, the temperatures at the hot and cold sides of the thermoelectric element, denoted by $T_1$ and $T_2$, differ from the reservoir temperatures (see Fig.\ref{TERmodel}). The temperatures are therefore ordered as 
\begin{equation}
T_{1} > T_{h} > T_{c} > T_{2}.
\end{equation}
Here, $(T_{1} - T_{h})$ and $(T_{c} - T_{2})$ represent temperature drops across the external thermal contacts, while $(T_1 - T_2)$ corresponds to the temperature difference across the thermoelectric material.

The heat fluxes at the hot and cold sides can be written as \cite{Pedersen2007}
\begin{equation}
\dot Q_h = \alpha T_1 I + \frac{1}{2} R I^2 - K_0 (T_1 - T_2),
\qquad
\dot Q_c = \alpha T_2 I - \frac{1}{2} R I^2 - K_0 (T_1 - T_2),
\label{qhc2}
\end{equation}
where $\alpha$ is the Seebeck coefficient, $R$ the internal electrical resistance, and $K_0$ the thermal conductance of the thermoelectric element. The first term represents Peltier heat, while the remaining terms describe irreversible contributions due to Joule heating and Fourier heat conduction within the TEM. 
The heat flux coming out of the cold reservoir ($\dot Q_c$) is 
also called the cooling power.
The coefficient of performance (COP) is defined as $\epsilon = \dot Q_c/P$, where $P = \dot Q_h - \dot Q_c$ is the input electrical power. 

\section{Exoreversible TER}
For reversible external thermal contacts,
the temperature gradients across the heat exchangers vanish,
i.e. $T_1 \to T_h$ and $T_2 \to T_c$,  so 
the heat flux equations are simplified to 
\begin{equation}
\dot Q_h = \alpha T_h I + \frac{1}{2} R I^2 - K_0 (T_h - T_c), 
\qquad
\dot Q_c = \alpha T_c I - \frac{1}{2} R I^2 - K_0 (T_h - T_c). 
\label{qcx}
\end{equation}

We introduce the dimensionless figure of merit for the 
refrigerator, $\mathtt{z} = \alpha^2 T_c/R K_0$.
Optimizing the cooling power in Eq. (\ref{qcx}) with respect to $I$ yields the optimal current as $I^{*}= {\alpha T_{c}}/{R}$.
The corresponding maximum cooling power and the COP are obtained as
\begin{equation}
    \dot{Q}_{c}^{*}=\frac{\alpha^{2} T_{c}^{2}}{2 R}-K_{0}\left(T_h-T_c \right),
    \qquad
    \epsilon_{\rm MCP}^{}= \frac{ \mathtt{z}T_c-2 (T_h-T_c)}{2 \mathtt{z}T_h}
\end{equation}
In the limit of very large $\mathtt{z}$ values,
we obtain \cite{Apertet2013} 
 \begin{equation}
      \epsilon_{\rm MCP}^{} = \frac{\epsilon_{\rm C}^{}}{2(1+\epsilon_{\rm C}^{})}, 
      \label{copexo}
 \end{equation}
 where $\epsilon_{\rm C}^{} = T_{c}/(T_{h} - T_{c})$
defines the Carnot COP 
of the refrigerator.
For $\epsilon_{\rm C}^{} \gg 1$, i.e. small temperature differences,  
$\epsilon_{\rm MCP}^{} \approx 1/2$.

\section{Endoreversible TER}
We may consider a TER operating under the highly idealized endoreversible assumption, where the only irreversibilities are from the finite-rate of heat exchange with the reservoirs owing to the finite thermal conductances at the hot and cold interfaces, while the  internal transport processes are assumed reversible ($R\to 0$ and $K_0 \to 0$), implying that $\mathtt{z} \to \infty$.
Though not a
realistic scenario, the analysis of this model serves 
to highlight the limiting cases of the more general
model, as we discuss in the next section.

With  the endoreversible approximation, 
the heat fluxes at the interfaces [Eqs. (\ref{qhc2})] are reduced to $\dot{Q}_h^{} = \alpha T_{1} I$ and $\dot{Q}_c^{} = \alpha T_{2} I$.
For simplicity, we consider the case of symmetric thermal conductances: $K_h = K_c = K$. Then, 
the heat fluxes between the heat reservoirs according to 
Newton's law, are: 
\begin{equation}
\dot{Q}_{h} = K_{}(T_{1}^{}-T_{h}^{}), 
\qquad
\dot{Q}_{c}  = -K_{}(T_{2}^{}-T_{c}^{}).
\label{qhcnew}
\end{equation}
The flux matching conditions 
yield the fluxes at the two interfaces as
\begin{equation}
\dot{Q}_{h}^{} = \frac{\alpha T_{h}^{}K_{}I}{K_{}-\alpha I}, 
	\qquad
	\dot{Q}_{c}^{} =  \frac{\alpha T_{c}^{}K_{}I}{K_{}+\alpha I}.
	\label{qhcn}
\end{equation}
Clearly, for a given current $I$, as the external conductance $K$
 becomes large, the thermal fluxes approach the
reversible expressions $\dot{Q}_{h}^{} \to \alpha T_{h} I$
and $\dot{Q}_{c}^{} \to \alpha T_{c} I$.
Now, it is the expression of $\dot{Q}_{c}$ in Eq. (\ref{qhcn}) which is not
optimizable w.r.t the variable $I$ (see Fig. 2). To explore
the effect of Newton's law close
to the reversible limit, i.e. assuming $K$ as large, but finite, 
we approximate the heat fluxes as
\begin{equation}
\dot{Q}_{h} = \alpha T_{h} I 
+ \frac{\alpha^2 T_{h}}{K} I^2, 
\qquad
\dot{Q}_{c} = \alpha T_{c} I - \frac{\alpha^2 T_{c}}{K} I^2 
\label{qhci2}.
\end{equation}
Apart from the reversible (Peltier) term, here is
an additional term due to finite thermal conductance of
the heat exchangers. This term is quadratic in the current $I$---a form similar to the heat flux in the exoreversible
model [Eq. (\ref{qcx})] which holds within the 
linear-irreversible framework \cite{Apertet_localtoglobal}. 
Thus, we may conclude that
the above approximation to the endoreversible model
is equivalent to the exoreversible
model with strong coupling ($K_0=0$) and that it also allows optimization of the cooling power. 

 \begin{figure}[h]
     \centering
     \includegraphics[width=0.6\linewidth]{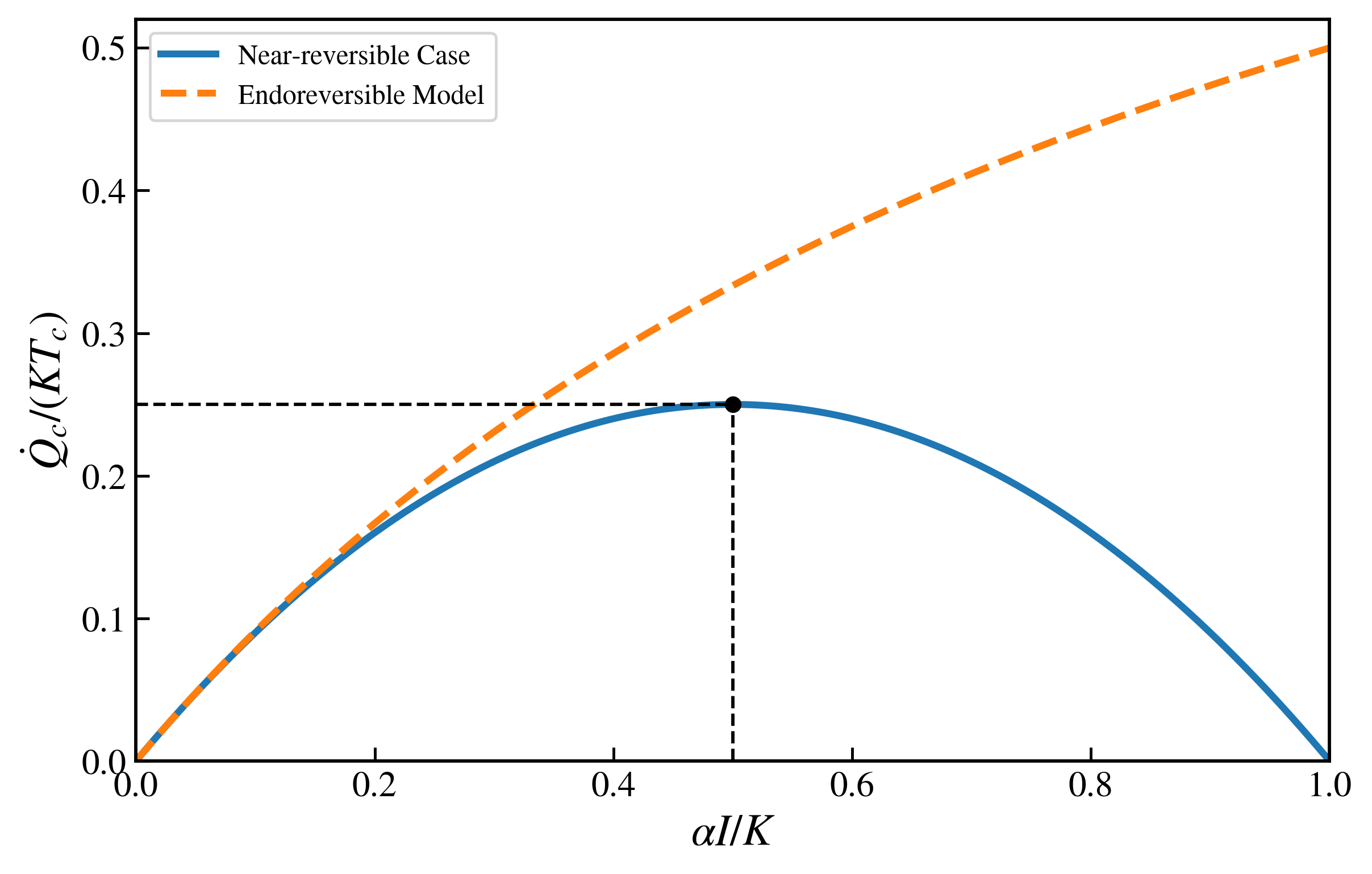}
\caption{Scaled cooling power in endoreversible
model with Newton's law [Eq. (\ref{qhcn})] 
increases monotonically with current $I$,
whereas in the near-reversible limit [Eq. (\ref{qhci2})], it displays a maximum at $\alpha I/K = 1/2$.}
\end{figure}

We now optimize the cooling power of Eq. (\ref{qhci2}) w.r.t. the current $I$. The optimal current and the maximum cooling power (MCP) are given by 
\begin{equation} 
I^{*} =
\frac{K}{2\alpha},
\label{ist}
\qquad
\dot{Q}_{c}^{}(I^*) =
\frac{K T_{c}}{4},
\end{equation}
which is also depicted in Fig. 2. 
The hot flux is $\dot{Q}_{h}^{}(I^*) =
3K T_{h}/4$. Then,
the COP is given by 
\begin{align}
\epsilon_{\rm MCP}^{} & =\frac{\epsilon_{\rm C}^{}}
{3+ 2 \epsilon_{\rm C}^{}}.
\label{coptr1}
\end{align}
Like the COP for exoreversible refrigerator 
[Eq. (\ref{copexo})],
the above expression also tends to the value  1/2 for
small temperature differences (see also Fig. 3).

\section{General case}
We place the above results in a more general context 
by considering both internal and 
external irreversibilities simultaneously. 
Combining the flux equations (\ref{qhc2}) for the TEM with Newtonion flux equations (\ref{qhcnew}), we can obtain explicit expressions for $T_1$ and $T_2$ \cite{Chakraborty2025}. Then, 
the explicit expressions for the heat fluxes 
[Eq. (\ref{qhc2})] are given by 
\begin{align}
\dot{Q}_{h} &= \frac{K \left(2K \alpha  T_h I  + ((K + 2 K_0)R + 2 \alpha^{2} T_h)  I^2 + \alpha R I^{3} - 2K K_0(T_h-T_c)\right)}{2(K(K+2K_0)-\alpha^2 I^2)}, \\
\dot{Q}_{c} &= \frac{K \left(2K \alpha  T_c I  - ((K + 2 K_0)R + 2 \alpha^{2} T_c)  I^2 + \alpha R I^{3} - 2K K_0(T_h-T_c)\right)}{2(K(K+2K_0)-\alpha^2 I^2)}.
\end{align}
However, analytic insight from the above  expressions is difficult,
so we resort to the regime of small external irreversibility,
as described below. 

\vspace{0.3cm}
\noindent
\textbf{Large-$K$ approximation}:
In the limit of large thermal conductance $K$,
i.e. small external irreversibility, we expand the heat fluxes up to first order in $1/K$. This yields
\begin{align}
\dot{Q}_{h} =& \left(1-\frac{2 K_0}{K}\right)
\left[\alpha T_h I - K_0(T_h-T_c)\right]
+ \left(\frac{R}{2}+\frac{\alpha^{2} T_{h}}{K}\right) I^{2}
+ \frac{\alpha R}{2 K} I^{3},
\label{hi4} \\
\dot{Q}_{c} =& \left(1-\frac{2 K_0}{K}\right)
\left[\alpha T_c I - K_0(T_h-T_c)\right]
- \left(\frac{R}{2}+\frac{\alpha^{2} T_{c}}{K}\right) I^{2}
+ \frac{\alpha R}{2 K} I^{3}.
\label{ci4}
\end{align}
The presence of $I^3$ term in the heat fluxes 
suggests that the above model is beyond the linear-response regime that was limited to a quadratic term in $I$, as discussed earlier. 
However, the input electrical power is given by
\begin{equation}
P= \left(1-\frac{2 K_0}{K}\right) \alpha (T_h-T_c) I
+\left(R+\frac{\alpha^{2}}{K}(T_{h}+T_{c})\right) I^{2}.
\label{Pwr}
\end{equation}
The quadratic expression for the power input is obtained
since the cubic terms cancel out. 

\vspace{0.3cm}
\noindent
\textbf{Optimization of cooling power}:
The cooling power $\dot{Q}_c$ [Eq. (\ref{ci4})] is maximized by
setting 
$\partial \dot{Q}_c / \partial I = 0$. The  optimal current is given by
\begin{equation}
I^*=\frac{K_0}{3\alpha k}\left(1+2 k \mathtt{z}
-\sqrt{1-2 k \mathtt{z}
\left(1-2 k(\mathtt{z}+3)\right)}\right).
\label{icc}
\end{equation}
Apart from the parameter $\mathtt{z}$,
the performance of the model also depends on the parameter $k = K_{0}/K$.
The condition $I^* > 0$ requires $k < 1/2$, and the second derivative test,
$\left.\partial^2 \dot{Q}_c / \partial I^2 \right|_{I=I^*} < 0$, confirms that this extremum corresponds to a maximum. The COP
$\epsilon^* = \dot{Q}_c (I^*)/P(I^*)$, 
is expressed in the form
\begin{equation}
\epsilon^* = \frac{\cal A}{\cal B},
\label{epab}
\end{equation}
where ${\cal A}$ and ${\cal B}$ are functions of $k$, $\mathtt{z}$, and $\epsilon_{\rm C}$:
\begin{align}
{\cal A} &= k \mathtt{z} \Big[\epsilon_{\rm C} \Big(4 a k (\mathtt{z}+3)-2 a
-4 k^2 \mathtt{z} (2 \mathtt{z}+9)
+6 k (\mathtt{z}-3)+3
-\frac{1-a}{k \mathtt{z}}\Big)
- 27 k (1-2 k)\Big], \nonumber\\
{\cal B} &= 3 (a-2 k \mathtt{z}-1)
\Big[\epsilon_{\rm C} (2 k \mathtt{z}+1) (a-2 k \mathtt{z}-1)
+ k \mathtt{z} (a-2 k (\mathtt{z}-3)-4)\Big],
\end{align}
with $a=\sqrt{1-2 k z (1-2 k (z+3))}$.

 \begin{figure}[h]
     \centering
     \includegraphics[width=0.6\linewidth]{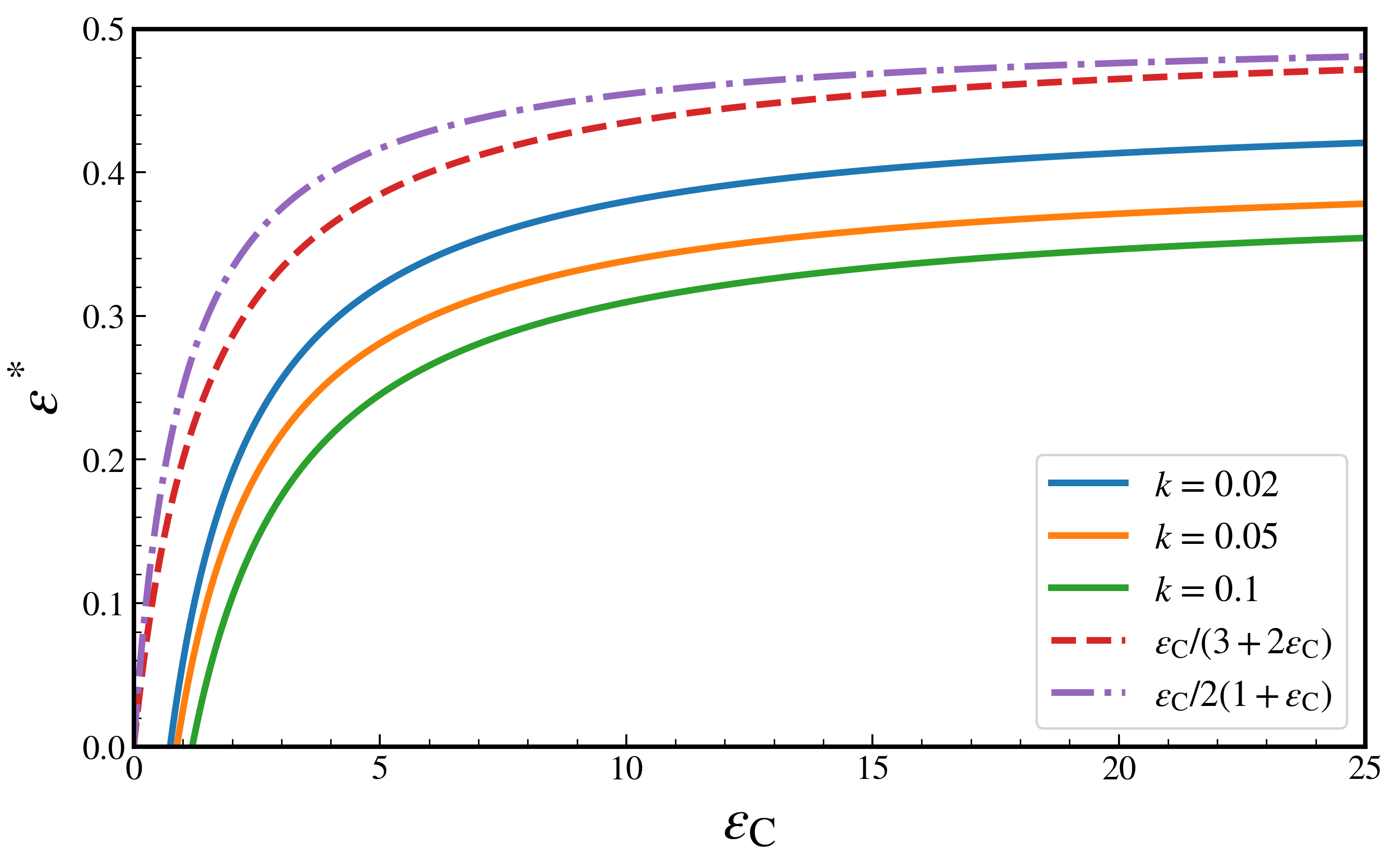}
\caption{The COP at maximum cooling power in different
approximations versus the Carnot COP ($\epsilon_{\rm C}^{}$). 
The top curve is for the exoreversible model with strong coupling,
followed by the symmetric endoreversible model close to the reversible limit. The lower three
curves display COP for the general case with both external and internal irreversibilities at different $k$ values,
while the figure of merit is kept $\mathtt{z}=3$.
}
\end{figure}

\vspace{0.3cm}
\noindent
\textbf{Limiting cases}:

(i) \textit{Endoreversible limit}:
Assuming $K_0 \to 0$ and $R \to 0$, 
Eq. (\ref{icc}) yields 
$I^* = K/2\alpha$, as found earlier [Eq. (\ref{ist})]. Also,
\begin{equation}
\dot{Q}_{c}^* = \frac{K T_{c}}{4}, \qquad
P^{*} = \frac{K (3T_{h}-T_{c})}{4}.
\end{equation}
The COP [Eq. (\ref{epab})] also reduces to
$\epsilon_{\rm C}/(3 + 2\epsilon_{\rm C})$ 
which was derived earlier within the symmetric, endoreversible
 approximation. 

(ii) \textit{Exoreversible limit}:
In the limit of reversible thermal contacts ($K \to \infty$),
we obtain, from Eq. (\ref{icc}), $I^* = \alpha T_c/R$. Then,
\begin{equation}
\lim_{k \to 0} \dot{Q}_c^* = K_0 T_c \left(\frac{\mathtt{z}}{2} -
\frac{1}{\epsilon_{\rm C}^{}} \right),
\qquad
\lim_{k \to 0} \epsilon^* =
\frac{\mathtt{z} \epsilon_{\rm C}^{}-2}{2 \mathtt{z}(1+\epsilon_{\rm C}^{})}.
\label{qc_epsk0}
\end{equation}
For strong coupling ($K_{0} \to 0$), $\mathtt{z}$ diverges and in this case COP yields the exoreversible limit [Eq. (\ref{copexo})].
Thus, we see that the general case at optimal cooling power  
unifies and  consistently recovers both the endoreversible and exoreversible limits as special cases.

We compare the expressions for COP at maximum 
cooling power for the different approximations in Fig. 3.
It is observed that the exoreversible model gives a higher
COP than the endoreversible model, both of them leading
to a value 1/2 for large $\epsilon_{\rm C}$ or 
small temperature differences. On the other hand, the 
presence of both internal and external irreversibilities
lowers the COP further. In this case, 
the limit of small temperature differences 
is dominated by the value:
\begin{equation}
     \epsilon^* \to \frac{1}{6} + \frac{\sqrt{1-2 k \mathtt{z}\left(1-2 k(\mathtt{z}+3)\right)}}{3(1+2k \mathtt{z})}.
\end{equation}
It is interesting to note
that the COP of actual thermoelectric coolers, say 
for single-stage applications, also have values in general
less than 1/2 \cite{Min2000, Gokcek2017}.

\section{Conclusions}
Both endoreversible and exoreversible models
are insufficient to 
simulate the performance of an actual TER. 
While the former completely neglects the 
intrinsic dissipative mechanisms within
the thermoelectric material, the latter
neglects the irreversibility in the 
external thermal couplings. 
In the refrigerator mode, the  
endoreversible model is further considered
to be problematic since 
the cooling power is not optimizable. 
We observe that if the endoreversible model 
with Newton's law is approximated 
in near-reversible regime, then it 
behaves analogous to the linear-irreversible 
model. Thereby, it becomes amenable to optimization
of the cooling power. 
We then considered the case of more realistic TER 
with both internal and external irreversibilities
with symmetric, large external conductances.  
The general expression of COP at 
optimal cooling power is found to be a function of
the system parameters such as figure of merit
and the ratio of internal to external thermal conductances,
apart from the ratio of reservoir temperatures. 
As desired, the optimized general case reduces 
to the  endoreversible and
exoreversible models in suitable limits.
The present approach thus broadens our understanding 
of these idealized cases within a general model,
and yields analytic estimates
to benchmark the performance of 
realistic thermoelectric devices.

\bibliography{mybib}

\end{document}